\documentclass{article}
\usepackage{spconf,amsmath,graphicx}
\usepackage{amsmath,amssymb,amsfonts}
\usepackage{mathrsfs}
\usepackage{amsthm}
\usepackage{rotating}
\usepackage{appendix}
\usepackage{ifpdf}
\usepackage[T1]{fontenc}
\usepackage{times}
\usepackage{newtxmath}
\usepackage{textcomp}%
\usepackage{xcolor}%
\usepackage{hyperref}
\usepackage{lipsum}
\usepackage[small,skip=3pt]{caption}
\usepackage{subcaption}
\usepackage{enumitem}
\usepackage{tikz}
\usepackage{comment}
\setlist{nosep, leftmargin=14pt}

\usepackage{mwe} 

\DeclareMathOperator*{\argmin}{arg\,min}
\newcommand{\txtb}[1]{\textcolor{black}{#1}}

\title{3D Image Super-Resolution by fluorophore fluctuations and MA-TIRF Microscopy reconstruction (3D-COL0RME)}
 \name{\begin{tabular}{c}Vasiliki Stergiopoulou$^{1}$, Luca Calatroni$^{1}$, Sébastien Schaub$^{2}$, and Laure Blanc-Féraud$^{1}$
 \end{tabular}}
\address{$^{1}$ Université Côte d’Azur, CNRS, INRIA, I3S, France \\
$^{2}$ Sorbonne Université, CNRS, LBDV, France \\}
\begin{document}
%
\maketitle
%

\begin{abstract}
We propose a 3D super-resolution approach to improve both lateral and axial spatial resolution \txtb{of a thin layer adjacent to the coverslip} in Total Internal \txtb{Reflection} Fluorescence (TIRF) imaging applications. Our approach, called 3D-COL0RME (3D - Covariance-based $\ell_0$ super-Resolution Microscopy with intensity Estimation) improves both lateral and axial resolution by combining sparsity-based modelling for precise molecule localisation and intensity estimation in the lateral plane with a 3D reconstruction procedure in the axial one using Multi-Angle TIRF (MA-TIRF). Differently from state-of-the-art approaches, 3D-COL0RME does not require specific fluorophores or specific illumination. We validate 3D-COL0RME on simulated MA-TIRF blinking-type data and on challenging real MA-TIRF acquisitions, showing significant resolution improvements.
\end{abstract}


\section{Introduction}
\label{sec:intro}
\vspace{-0.1cm}

Fluorescence microscopy is an imaging technique which allows the real-time observation of sub-cellular entities in live samples. Due to light diffraction, however, the spatial resolution normally achieved by means of light microscopy techniques is limited. In standard acquisition settings, the diffraction limit, i.e. the size of the biological entities we are able to distinguish, is approximately equal to $200 nm$ in the lateral plane and to $500 nm$ in the axial one. Several biological entities of interest, however, have size smaller than these values. The implementation of appropriate super-resolution techniques can thus significantly improve the visualisations of invisible sub-cellular entities, by allowing the understanding of the biological functions happening at the molecular level. 

There are many super-resolution techniques appeared in the literature. Some of them are able to gain resolution in 3D. The widely known are 3D-STORM \cite{3D_STORM} and 3D-StED \cite{3DSTED} that can achieve very good levels of spatial resolution but at the expense of long acquisition times (especially for StORM), fast photobleaching and they require special fluorophores (especially for StED). The most resolutive method in 3D, to our knowledge, is iPALM \cite{iPALM} which combines StORM approach and interferometry at the price of a highly complex prototype. On the other side 3D SIM \cite{3D_SIM} has short acquisition times, so improved temporal resolution, but limited spatial resolution.

In this work, we relax the requirement of special equipment and propose a super-resolution method for live-cell imaging with fine spatio-temporal resolution, which exploits the stochastic temporal fluctuations of \emph{standard} fluorophores. The diffraction-limited images we aim to process are acquired by a Multi-Angle Total Internal Reflection Fluorescent (MA-TIRF) microscope that benefits from low photobleaching, low photodamage and can excite a broad spectrum of common fluorophores. \txtb{However, due to the fast decay of the evanescent field produced by total internal reflection, TIRF imaging is limited to a thin layer (500-800 nm) adjacent to the coverslip, as only the fluorophores in this layer are excited.} 

Compared to standard TIRF microscopy acquisitions \cite{TIRF_axelrod81,AXELROD2008169}, MA-TIRF further provides depth information as it allows the sample's optical sectioning at different axial levels according to the chosen incident angle of the illumination. Given a stack of MA-TIRF acquisitions, a 3D super-resolved image can be estimated by means of suitable super-resolution methods. For improving the spatial resolution in the lateral plane, we consider in this work the Covariance-based $\ell_0$ super-Resolution Microscopy approach with intensity Estimation (COL0RME) \cite{stergiopoulou_ISBI,stergiopoulou_BioIm}, while for the axial plane, we consider the MA-TIRF reconstruction algorithm proposed in \cite{soubies_journal,soubies_ISBI}. Our combined method is referred to as 3D-COL0RME.

Our paper is structured as follows: in Section \ref{model}, we describe in mathematical terms the 3D MA-TIRF super-resolution problem. In Section \ref{sec:3D-COL0RME} we discuss the COL0RME and MA-TIRF reconstruction models allowing for improved resolution in the lateral and axial plane, respectively. Finally, in Section \ref{sec:results} we report results computed both on simulated and real data.

\section{Problem formulation}
\label{model}
\vspace{-0.1cm}
To start with, TIRF microscopy exploits the properties of the evanescent field appearing in a small region adjacent to the interface between two mediums with different refractive indices. When the light propagates from a medium with a high refractive index to a medium with a lower refractive index, there is a critical angle ($\alpha_c$) beyond which the total amount of light is reflected. For angles $\alpha > \alpha_c$ the light beam is totally reflected, so the electromagnetic field (evanescent wave) penetrates into the medium with an intensity $I(z,\alpha)$ that exponentially decays in the axial direction $z$ and is given by:
\begin{equation}\label{I}
    \textstyle
    I(z,\alpha)= I_0 (\alpha) e^{-zp(\alpha)}, \quad p(\alpha) = \gamma \sqrt{(\sin^2(\alpha) - \sin^2 (a_c))}
\end{equation}
where $p(\alpha)$ is the inverse of the penetration depth and $\gamma$ is a parameter that depends on the index of the incident medium and the excitation wavelength. See \cite{TIRF_axelrod81, AXELROD2008169} for more details.

Given a set of images $\{\mathbf{x}_{\alpha_q} \in \mathbb{R}^{{L^2}}\}_{q=1}^{N_q}$ obtained with different incident angles of the illumination beam $\{\alpha_q > \alpha_c\}_{q=1}^{N_q} $, we are able to retrieve the elevation in the axial direction ($z$) in each pixel  $\mathbf{f}\in\mathbb{R}^{{L^2} \times N_z}$, by considering the following problem:
\begin{equation}\label{model_MATIRF}
    \textstyle
    \text{find }\quad \mathbf{f}\in\mathbb{R}^{{L^2} \times N_z}\quad\text{s.t.} \quad   \mathbf{x}_{\alpha_q} = \mathbf{W}_{\alpha_q} \mathbf{f},
\end{equation} 
where $\mathbf{W}_{\alpha_q}:\mathbb{R}^{{L^2} \times N_z} \rightarrow \mathbb{R}^{L^2}$ is an operator representing the weighted summation of the $N_z$ slices of $\mathbf{f}$, with weights related to the angle $\alpha_q$ and equal to the intensity factors $\{I(z,\alpha_q)\}_{z=1}^{N_z}$ of the evanescent field.

Due to light diffraction, the resolution of the acquisitions is still limited in the lateral plane. Therefore, we propose the computation of highly-resolved images in the lateral plane for each one of the incident angles $\alpha_q$. By acquiring a sequence of $T$ images $\{\mathbf{g}_{\alpha_q,t} \in \mathbb{R}^{M^2}\}_{t=1}^T$, for each angle $\alpha_q$, and by exploiting the independence of the random fluctuations of individual fluorescent molecules, we aim to find $\mathbf{x}_{\alpha_q} \in \mathbb{R}^{{L^2}}$, the laterally super-resolved image defined on a $r$-times finer grid, if $L=rM$, $r>1$. Mathematically, we can write:
\begin{equation}\label{model_COL0RME}
    \textstyle
    \mathbf{x}_{\alpha_q} :=\frac{1}{T}\sum\limits_{t=1}^T\mathbf{u}_{\alpha_q,t}\quad\text{with}\quad \mathbf{g}_{\alpha_q,t}  = \mathbf{S} \mathbf{H} \mathbf{u}_{\alpha_q,t} + \mathbf{b}_{\alpha_q} + \mathbf{n}_{\alpha_q,t},
\end{equation}
where $\mathbf{S}:\mathbb{R}^{L^2} \to \mathbb{R}^{M^2}$ is a down-sampling operator summing every $r$ consecutive pixels in both dimensions and $\mathbf{H}:\mathbb{R}^{L^2} \to \mathbb{R}^{L^2}$ is a convolution operator defined by the point-spread function (PSF) of the optical imaging system. By $\mathbf{b}_{\alpha_q} \in \mathbb{R}^{M^2}$ we denote the stationary background  for each incident angle $\alpha_q$ and by $\mathbf{n}_{\alpha_q,t} \in \mathbb{R}^{M^2}$ a vector of independent and identically distributed (i.i.d) Gaussian entries of zero mean and constant variance $s \in \mathbb{R}_+$ modelling the presence of electronic noise. For simplicity, we consider in the model only additive Gaussian noise although, in practice, signal-dependent Poisson noise should be considered. This Poisson noise is however present in simulated data in Section \ref{sec:results}. 

In the modelling above, the desired 3D super-resolved image  $\mathbf{f} \in \mathbb{R}^{{L^2} \times N_z}$  is thus estimated given a stack of  blinking acquisitions $ \{ \{\mathbf{g}_{\alpha_q, t}\}_{t=1}^T \}_{q=1}^{N_q}$ acquired at different angles $\alpha_q>a_c, \forall q\in\left\{1,\dots,N_q\right\}$.

\section{3D-COL0RME}
\label{sec:3D-COL0RME}
\vspace{-0.1cm}

We describe in this section how the two ill-posed inverse problems \eqref{model_MATIRF} and \eqref{model_COL0RME} can be solved in a sequential way by means of appropriate sparse regularisation models. Note that by solving \eqref{model_COL0RME} a super-resolved image $\hat{\mathbf{x}}_{\alpha_q}$ can be estimated for each incident angle $\alpha_q$ of the illumination beam. Those images serve as input data for problem  \eqref{model_MATIRF} where the objective consists in finding the desired 3D image $\mathbf{f}$ with improved spatial and axial resolution. 

For the resolution of \eqref{model_COL0RME} we  use the COL0RME approach  \cite{stergiopoulou_ISBI,stergiopoulou_BioIm} which is based on the formulation of a two-step procedure which, by solving suitable sparse optimisation problems, computes an accurate estimation of both sample support and intensity. For solving \eqref{model_MATIRF} we use the 3D MA-TIRF reconstruction algorithm proposed in previous works \cite{soubies_journal,soubies_ISBI}. 

\subsection{Support and intensity estimation via COL0RME}
\label{sec:COL0RME}

The method COL0RME \cite{stergiopoulou_ISBI, stergiopoulou_BioIm} is composed of two steps: a support estimation step and an intensity estimation one. The two steps are performed sequentially.
The main idea of COL0RME consists in exploiting the temporal and spatial independence of the fluorescent emitters  by  a sparse approximation of their second-order statistics. A formulation of problem \eqref{model_COL0RME} in the covariance domain is the following (see \cite{stergiopoulou_ISBI, stergiopoulou_BioIm} for more details):
\begin{equation} \label{prob:a_q}
    \textstyle
    \mathbf{r^{\alpha_q}_g} = (\mathbf{\Psi} \odot \mathbf{\Psi}) \mathbf{r^{\alpha_q}_u} + s^{\alpha_q} \mathbf{v_I},
\end{equation}
where, for all angles $\alpha_q$, $\mathbf{r^{\alpha_q}_g} \in \mathbb{R}^{M^4}$ is the vectorised form of the covariance matrix of the raw data $\{\mathbf{g}_{\alpha_q,t}\}_{t=1}^T$, $\mathbf{r^{\alpha_q}_u}\in\mathbb{R}^{L^2}$ is the vector of the auto-covariances of the high-resolution images $\{\mathbf{u}_{\alpha_q,t}\}_{t=1}^T$, $s^{\alpha_q}\in\mathbb{R}_+$ is the (unknown) Gaussian noise variance, $\mathbf{\Psi}:=\textbf{SH}\in\mathbb{R}^{M^2 \times L^2}$, $\odot$ denotes the column-wise Kronecker product and $\mathbf{v_I} \in \mathbb{R}^{M^4}$ is the vectorised form of the identity matrix  $\mathbf{I_{M^2}}$. 

Note that due to the MA-TIRF setup, the support of the entire sample can be found by solving \eqref{prob:a_q} only in correspondence with the angle $\tilde{\alpha}$ closest to the critical one $\alpha_c$, as it will contain information of molecules located in the whole depth of investigation. This corresponds to a significant computational gain as \eqref{prob:a_q} needs thus to be solved only once. By denoting with $\tilde{\mathbf{r}}_\mathbf{g}$ the covariance matrix associated to $\{\mathbf{g}_{\tilde{\alpha},t}\}_{t=1}^T$, then the following problem can be considered for solving \eqref{prob:a_q}:
{\begin{equation}\label{support_est}
    \textstyle
    ({\hat{\mathbf{r}}_\mathbf{u}}, \hat{s}) \in \argmin\limits_{\mathbf{\tilde{r}_u} \in \mathbb{R}_+^{L^2},~ \tilde{s}\in\mathbb{R}+} \left( \frac12 \| \tilde{\mathbf{r}}_\mathbf{g} -(\mathbf\Psi \odot \mathbf\Psi) \mathbf{\tilde{r}_u} - \tilde{s} \mathbf{v_I} \|_2^2+ {\cal{R}}(\mathbf{\tilde{r}_u};\lambda)\right),
\end{equation}}
where $\lambda>0$ is a regularization parameter, $\cal{R}(\cdot;\lambda)$ is a sparsity-promoting penalty, that could be either the $\ell_1$-norm or the continuous exact relaxations of the $\ell_0$ (CEL0) \cite{CELO} penalty. Numerically, problem \eqref{support_est} can be solved by alternate minimisation and by means of standard sparse optimisation solvers (such as FISTA, iterative reweighted $\ell_1$\ldots), see, e.g., \cite{chambolle_pock_2016} for a review.

Given $\hat{\mathbf{r}}_\mathbf{u}$, the support of interest $\Omega$ is simply $\Omega = \left\{i: (\hat{\mathbf{r}}_{\mathbf{u}})_i\neq 0\right\}=\left\{i: (\hat{\mathbf{x}}_{\alpha_q})_i\neq 0\right\}$. By an accurate estimation of the noise variance $\hat{s}$ and by means of the discrepancy principle \cite{DiscInvProb_Hansen}, an automatic intensity estimation procedure can be now designed as a second step of COL0RME. In our scenario, for each angle  $\alpha_q$, $q\in\left\{1,...,N_q\right\}$, the mean intensity image $\hat{\mathbf{x}}_{\alpha_q}$ restricted to the estimated support $\Omega$ and the smoothly varying background $\hat{\mathbf{b}}_{\alpha_q}$can be estimated from the empirical temporal mean of the acquired stack $\{\overline{\mathbf{g}}_{\alpha_q}\}_{q=1}^{N_q}$ by solving:
\begin{align}\label{intensity_est}
    \textstyle
    (\hat{\mathbf{x}}_{\alpha_q},\hat{\mathbf{b}}_{\alpha_q}) \in &\argmin\limits_{\mathbf{x}\in\mathbb{R}_+^{|\Omega|},~ \mathbf{b}\in\mathbb{R}_+^{M^2}} \bigg( \frac12 \|\mathbf{\Psi_\Omega} \mathbf{x} - (\overline{\mathbf{g}}_{\alpha_q} - \mathbf{b})\|_2^2\\
    &\quad+ \frac{\mu}{2} \|\nabla_{\Omega}\mathbf{x}\|_2^2 + \frac{\beta}{2} \|\nabla\mathbf{b}\|_2^2 \bigg), \quad \forall q\in\left\{1,\dots,N_q\right\}. \notag
\end{align}
Here, the parameter $\mu$ can be automatically estimated via discrepancy principle while $\beta>0$ does not require very fine tuning.
By $\mathbf{\nabla}\in\mathbb{R}^{2 M^2\times M^2}$ we denote the discrete gradient operator, while $\mathbf{\Psi_\Omega} \in\mathbb{R}^{M^2\times |\Omega|}$ is a matrix whose $i$-th column is extracted from $\mathbf\Psi$ for all indexes $i \in \Omega$ and $\mathbf{\nabla}_\Omega$ is the discrete gradient operator restricted to $\Omega$.  Problem \eqref{intensity_est} can be solved efficiently via (proximal) gradient-type algorithms.


\subsection{MA-TIRF reconstruction}
\label{sec:MA-TIRF}

Having all the estimated COL0RME images $\hat{\mathbf{x}}_{\alpha_q}, \forall q=\{1,\ldots,N_q\}$ at hand, we can use them to solve the problem \eqref{model_MATIRF}.
To estimate the 3D super-resolved image $\hat{\mathbf{f}}\in \mathbb{R}^{{L^2} \times N_z}$, we thus follow \cite{soubies_journal,soubies_ISBI} and look for solutions of
\begin{equation} \label{MA-TIRF_reconst}
    \textstyle
   \hat{\mathbf{f}} \in \argmin\limits_{\mathbf{f}\in \mathbb{R}_+^{{L^2}\times N_z}}\left( \sum\limits_{q=1}^{N_q} \frac12\|\mathbf{W}_{\alpha_q} \mathbf{f} - \hat{\mathbf{x}}_{\alpha_q}\|_2^2 + \kappa {\cal{R}}(\nabla\mathbf{f}))\right),
\end{equation}
where $\kappa>0$, $\{\hat{\mathbf{x}}_{\alpha_q}\}_{q=1}^{N_q}$ are the super-resolved COL0RME images and $\mathbf{W}_{\alpha_q}:\mathbb{R}^{{L^2} \times N_z} \rightarrow \mathbb{R}^{L^2}$ is the discrete TIRF operator related to the angle $\alpha_q$, while $\cal{R}(\nabla\cdot)$ can be the Hessian Shatten-norm or the Total-Variation (TV) regularization, depending on whether Hessian- or gradient-sparsity is seeked, respectively.

\section{Results}
\label{sec:results}
\begin{figure}
    \centering
     \begin{subfigure}[b]{0.17\textwidth}
         \centering
         \includegraphics[trim=40 0 40 0,clip,width=\textwidth]{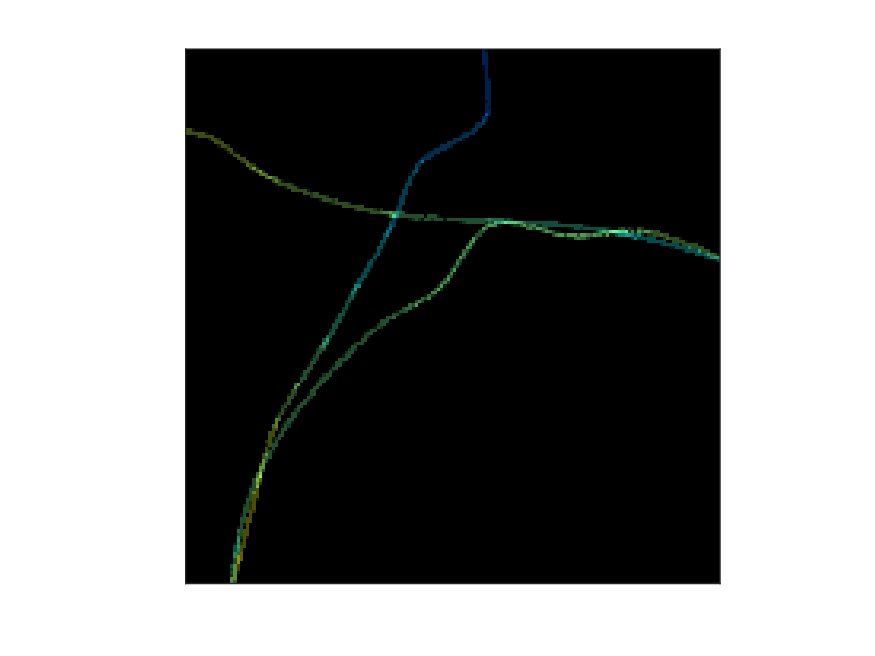}
         \caption{}
          \label{fig: GT}
     \end{subfigure}
    \begin{subfigure}[b]{0.03\textwidth}
         \centering
         \includegraphics[trim=180 0 180 0,clip,width=\textwidth]{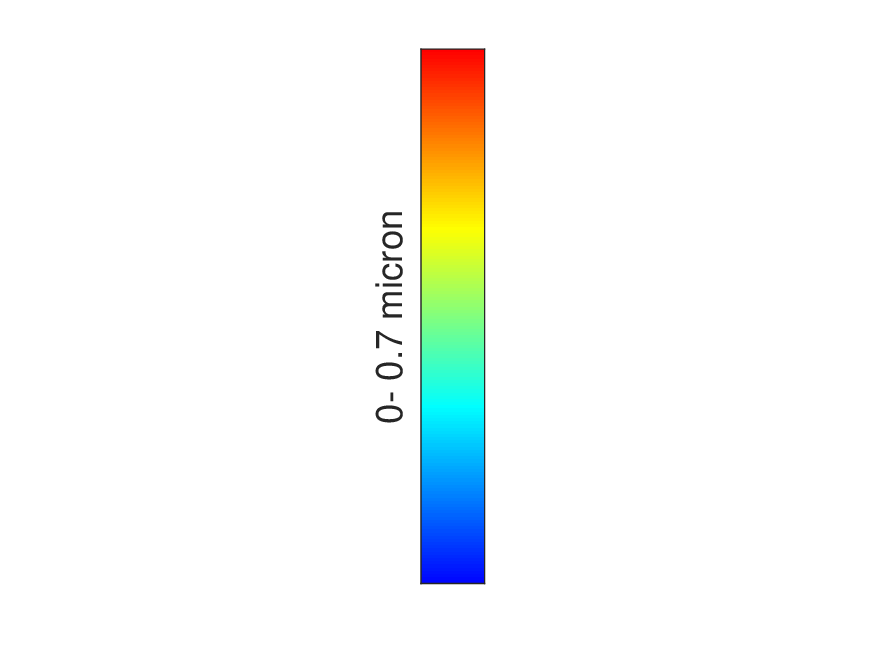}
         \caption*{}
     \end{subfigure}
     \hspace{0.5cm}
     \begin{subfigure}[b]{0.17\textwidth}
         \centering     
         \includegraphics[trim=40 0 40 0,clip,width=\textwidth]{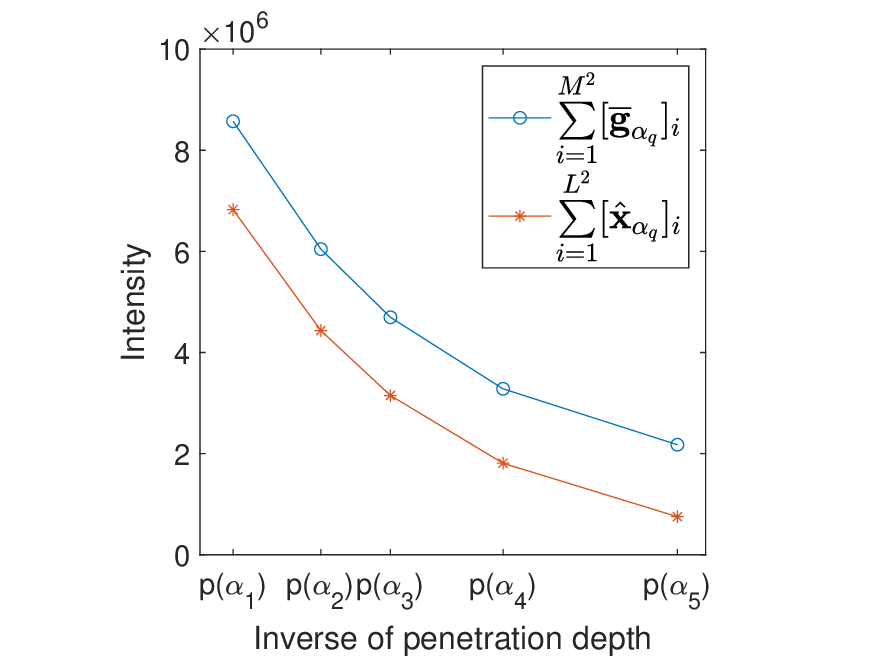}
         \caption{}
     \end{subfigure}
      \caption{(a) Ground truth tubulin image with depth information, \\
      (b) The exponential decay of the global intensity of the diffraction limited and the super-resolved COL0RME images with respect to the inverse of the penetration depth $\{p(\alpha_q)\}_{q=1}^{5}$, see \eqref{I} .}
\vspace{-0.4cm}
\end{figure}

\vspace{-0.1cm}
\textbf{Simulated MA-TIRF SOFI data.} We start by applying 3D-COL0RME to process simulated 3D tubulin images.
To simulate the data, we first set the 3D spatial pattern (see Figure \ref{fig: GT}) using the SMLM 2016 MT0 microtubules dataset \footnote{\href{http://bigwww.epfl.ch/smlm/datasets/index.html}{http://bigwww.epfl.ch/smlm/datasets/index.html}}; temporal fluctuations are simulated by using the SOFI simulation tool \cite{SOFItool} upon a specific choice of parameters (see below). For five different angles $\{\alpha_q\}_{q=1}^5$ of the illumination beam, with $\alpha_c < \alpha_1<\dots<\alpha_5$,  a stack of 500 frames is simulated. The fluctuations' parameters are chosen as: $20$ms for on-state average lifetime, $40$ms for off-state average lifetime, $35$s for average time until bleaching (so that little bleaching, around 18\% is practically observed
) and frame rate of $100$ frames per second (fps). The PSF used has a full-width-half-maximum (FWHM) of approximately $229$nm while the pixel size is chosen to be equal to $100$nm. Spatially varying background was added to the acquisition as well as additive Gaussian noise of signal-to-noise ration (SNR) equal to $14.75$ dB.

\begin{figure}[h]
     \centering
     \begin{subfigure}[b]{0.067\textwidth}
         \centering
         \includegraphics[width=\textwidth]{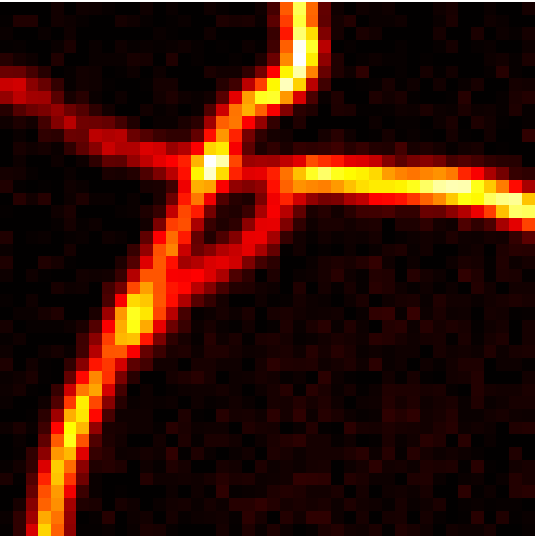}
     \end{subfigure}
     \hfill
      \begin{subfigure}[b]{0.067\textwidth}
         \centering
         \includegraphics[width=\textwidth]{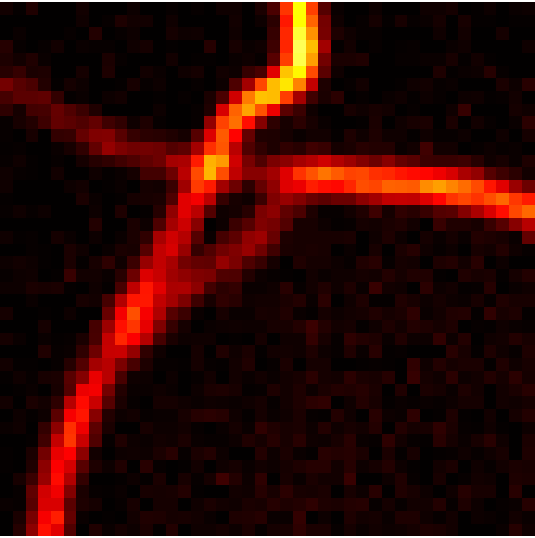}
     \end{subfigure}
     \hfill
      \begin{subfigure}[b]{0.067\textwidth}
         \centering
         \includegraphics[width=\textwidth]{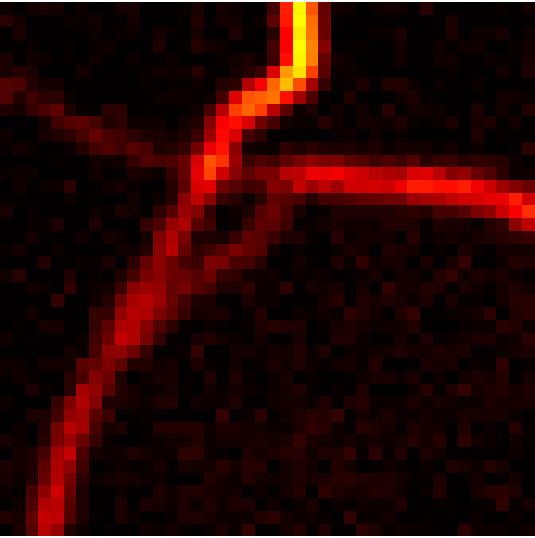}
     \end{subfigure}
     \hfill
      \begin{subfigure}[b]{0.067\textwidth}
         \centering
         \includegraphics[width=\textwidth]{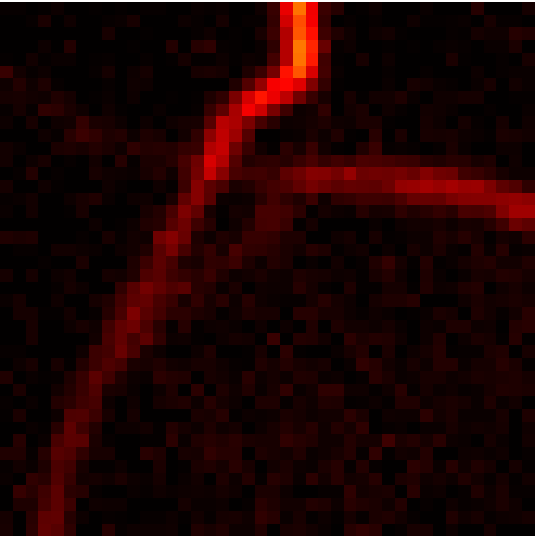}
     \end{subfigure}
     \hfill
      \begin{subfigure}[b]{0.067\textwidth}
         \centering
         \includegraphics[width=\textwidth]{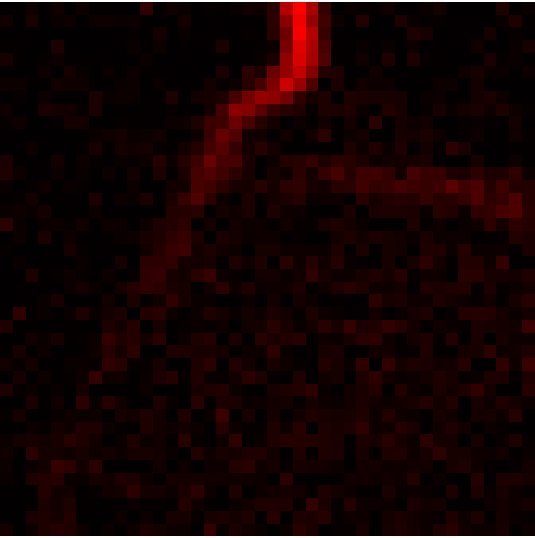}
     \end{subfigure}
     \hfill
      \begin{subfigure}[b]{0.016\textwidth}
         \centering
         \includegraphics[width=\textwidth]{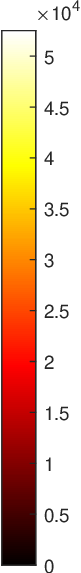}
     \end{subfigure}
     
     \vspace{0.05cm}

          \begin{subfigure}[b]{0.067\textwidth}
         \centering
         \includegraphics[width=\textwidth]{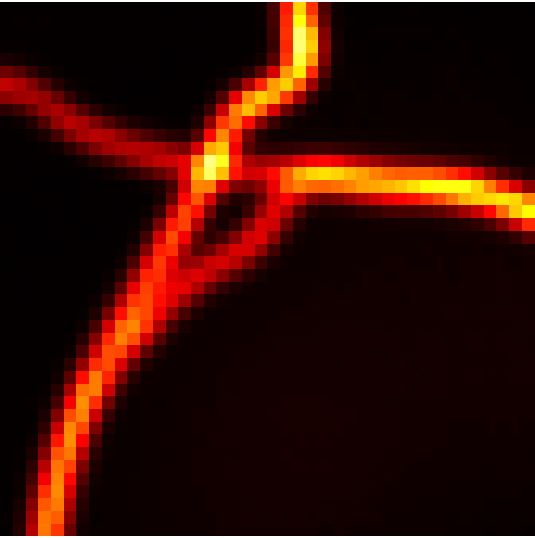}
     \end{subfigure}
     \hfill
      \begin{subfigure}[b]{0.067\textwidth}
         \centering
         \includegraphics[width=\textwidth]{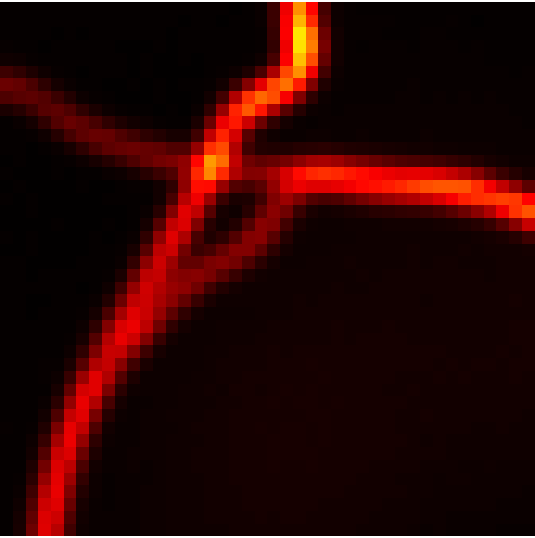}
     \end{subfigure}
     \hfill
      \begin{subfigure}[b]{0.067\textwidth}
         \centering
         \includegraphics[width=\textwidth]{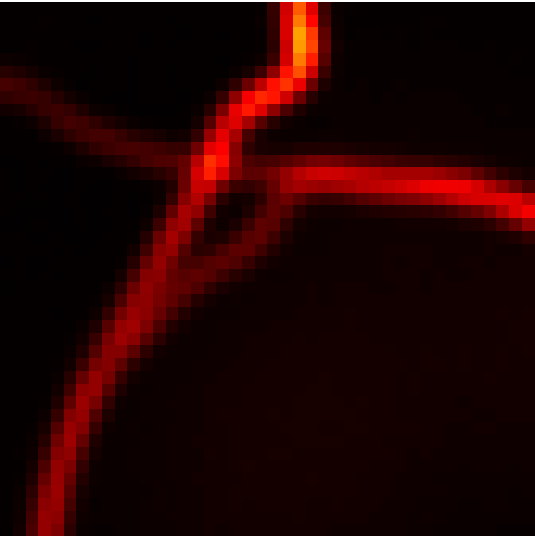}
     \end{subfigure}
     \hfill
      \begin{subfigure}[b]{0.067\textwidth}
         \centering
         \includegraphics[width=\textwidth]{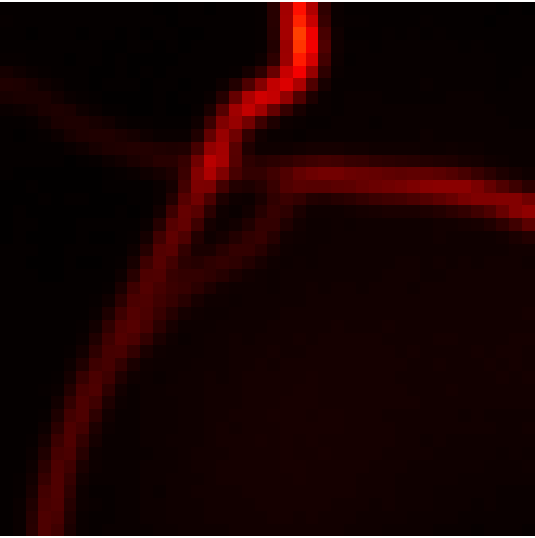}
     \end{subfigure}
     \hfill
      \begin{subfigure}[b]{0.067\textwidth}
         \centering
         \includegraphics[width=\textwidth]{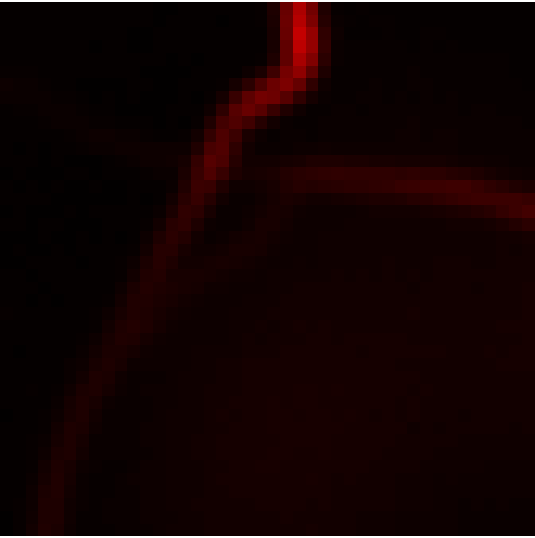}
     \end{subfigure}
     \hfill
      \begin{subfigure}[b]{0.016\textwidth}
         \centering
         \includegraphics[width=\textwidth]{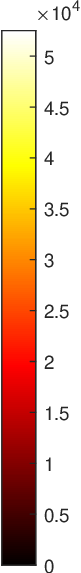}
     \end{subfigure}
     
     \vspace{0.05cm}
     
          \begin{subfigure}[b]{0.067\textwidth}
         \centering
         \includegraphics[width=\textwidth]{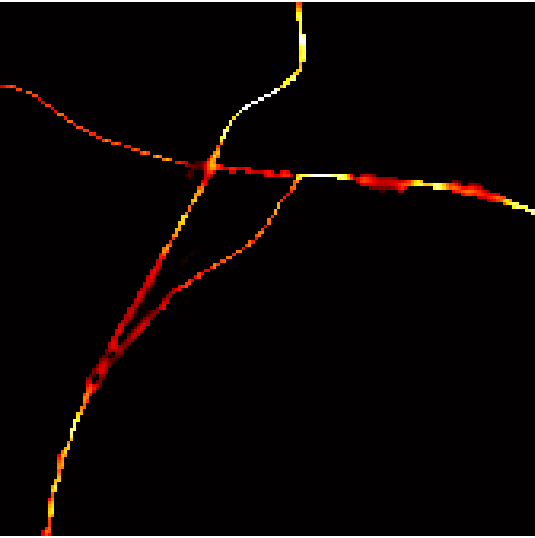}
         \caption*{$\alpha_1$}
     \end{subfigure}
     \hfill
      \begin{subfigure}[b]{0.067\textwidth}
         \centering
         \includegraphics[width=\textwidth]{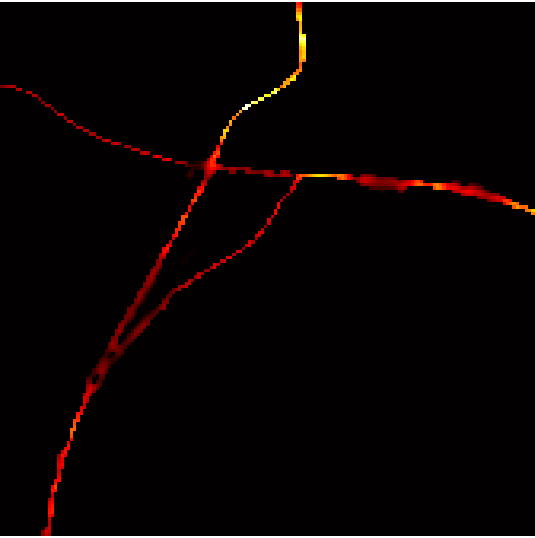}
          \caption*{$\alpha_2$}
     \end{subfigure}
     \hfill
      \begin{subfigure}[b]{0.067\textwidth}
         \centering
         \includegraphics[width=\textwidth]{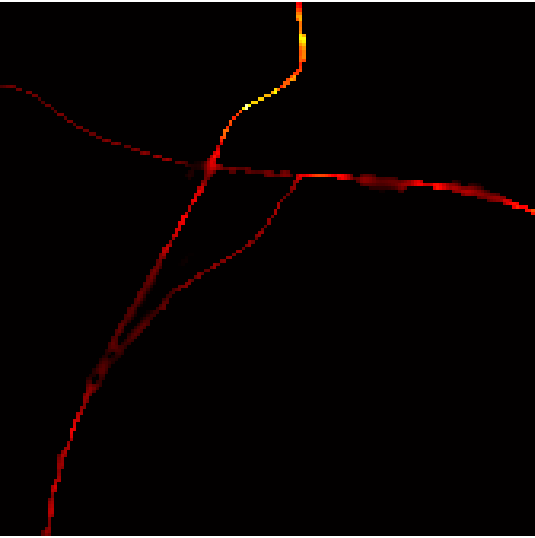}
         \caption*{$\alpha_3$}
     \end{subfigure}
     \hfill
      \begin{subfigure}[b]{0.067\textwidth}
         \centering
         \includegraphics[width=\textwidth]{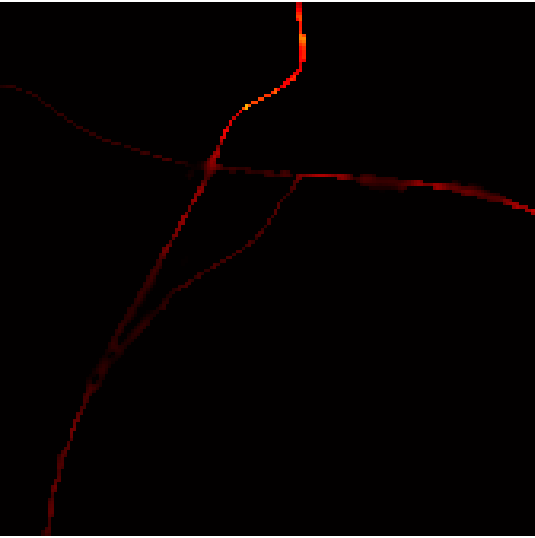}
         \caption*{$\alpha_4$}
     \end{subfigure}
     \hfill
      \begin{subfigure}[b]{0.067\textwidth}
         \centering
         \includegraphics[width=\textwidth]{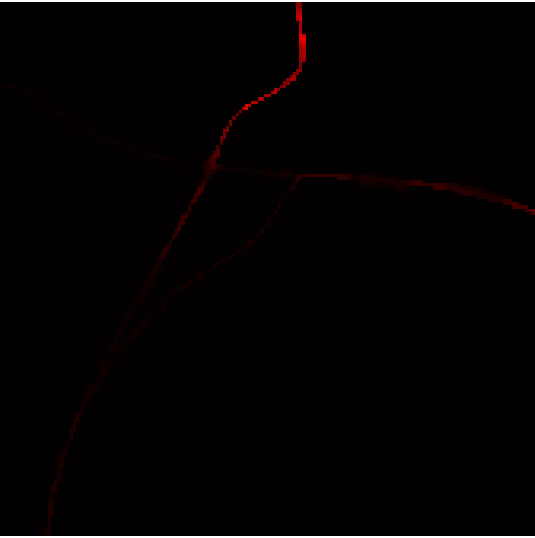}
         \caption*{$\alpha_5$}
     \end{subfigure}
     \hfill
      \begin{subfigure}[b]{0.016\textwidth}
         \centering
         \includegraphics[width=\textwidth]{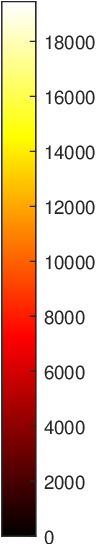}
      \caption*{}
     \end{subfigure}
     \caption{For each $\{\alpha_q\}$, $q=1,\ldots,5$: (first row) one frame of the acquired stack $\mathbf{g}_{\alpha_q,t}$, (second row) the temporal mean $\bar{\textbf{g}}_{\alpha_q}$, (third line) 2D COL0RME results $\hat{\mathbf{x}}_{\alpha_q}$.}
    \label{fig: simulated data}
\end{figure}

\begin{figure}
    \vspace{-0.2cm}
    \centering
    \begin{subfigure}[b]{0.18\textwidth}
         \centering
         \includegraphics[trim=40 0 40 0,clip,width=\textwidth]{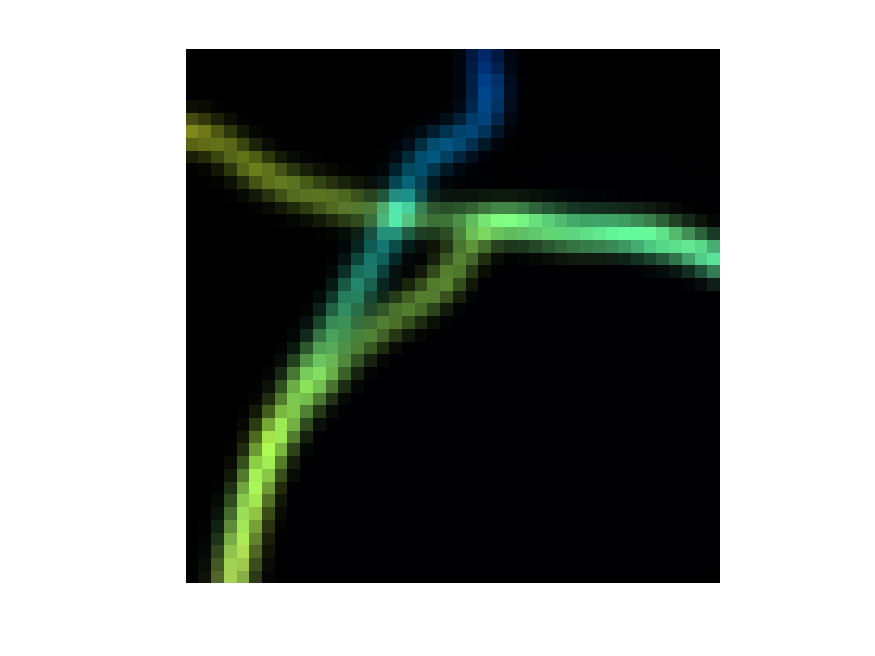}
         \vspace{-2\baselineskip}
         \caption{}
         \label{MA-TIRF}
     \end{subfigure}
     \hfill
      \begin{subfigure}[b]{0.18\textwidth}
         \centering
         \includegraphics[trim=40 0 40 0,clip,width=\textwidth]{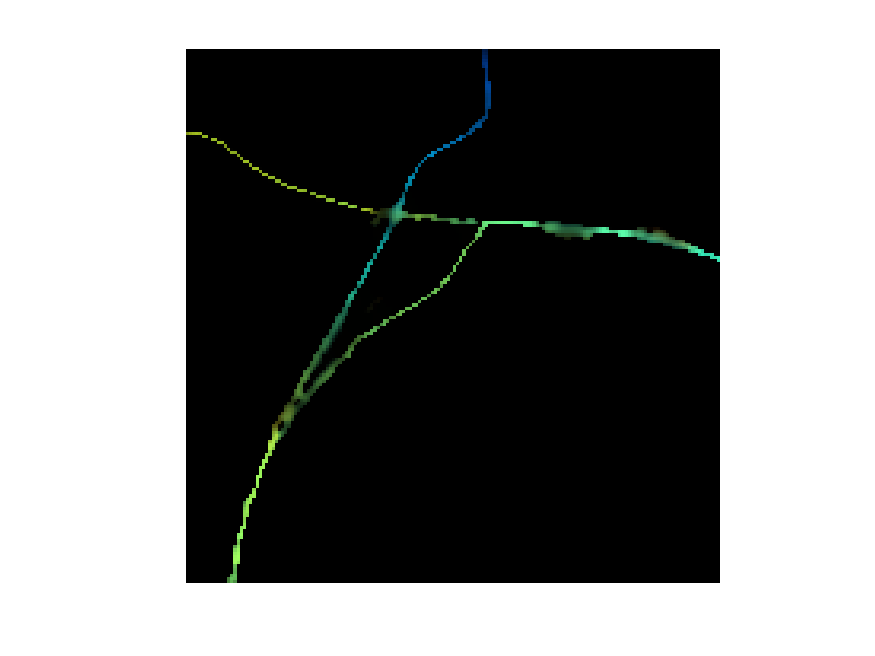}
        \vspace{-2\baselineskip}
          \caption{}
         \label{3D-COL0RME}
     \end{subfigure}
      \hfill
      \begin{subfigure}[b]{0.03\textwidth}
         \centering
         \includegraphics[trim=180 32 180 0,clip,width=\textwidth]{images/5angles_3Dcolorbar.eps}
         \caption*{}
     \end{subfigure}
     \caption{(a) MA-TIRF reconstruction result (b) Super-resolved 3D-COL0RME reconstruction. Colour quantifies sample depth.}
     \vspace{-0.4cm}
\end{figure}

In Figure \ref{fig: simulated data}, a single frame of the acquired stack as well as its temporal mean (diffraction-limited image) is shown for each incident angle of the illumination beam in the first and second line, respectively. In the third line, the super-resolved images $\{\hat{\mathbf{x}}_{\alpha_q}\}_{q=1}^5$ computed by solving COL0RME models \eqref{support_est}-\eqref{intensity_est} at each angle with a super-resolution factor of $4$. Finally, in Figure \ref{3D-COL0RME} we show the 3D MA-TIRF reconstruction result obtained by solving \eqref{MA-TIRF_reconst} using $\{\hat{\mathbf{x}}_{\alpha_q}\}_{q=1}^5$ as input. The regularization parameters that do not allow automatic estimation have been chosen empirically, \txtb{while $\ell_1$- and TV-regularization has been used for the problems \eqref{support_est} and \eqref{MA-TIRF_reconst} respectively.}

For comparison, we further plot in Figure \ref{MA-TIRF} the result of the MA-TIRF approach without deconvolution previously considered \cite{soubies_journal,soubies_ISBI}: we can clearly observe that, compared to 3D-COL0RME, shows fine axial, but poor spatial resolution.

\textbf{Real MA-TIRF data}. We further apply 3D-COL0RME on a dataset of images \txtb{of tubulins
stained with Alexa Fluor 488 that are} acquired by a real MA-TIRF microscope. A sequence of $T=500$ frames is processed for each $\alpha_q, q=0,\ldots,4$. As $\alpha_0 < \alpha_c $, no propagation of the evanescent wave is observed for it. However, such angle is used only for a more precise support estimation. Namely, as a final support we consider the superposition of the supports estimated for angles $\alpha_0$ and $\alpha_1$. The total acquisition time of the whole dataset is approximately $2$min, the pixel size of the CCD camera used is 106nm and the FWHM of the PSF has been measured experimentally and is equal to $292$nm.

\begin{figure}[h]
    \vspace{-0.2cm}
     \centering
     \begin{subfigure}[b]{0.08\textwidth}
         \centering
         \includegraphics[width=\textwidth]{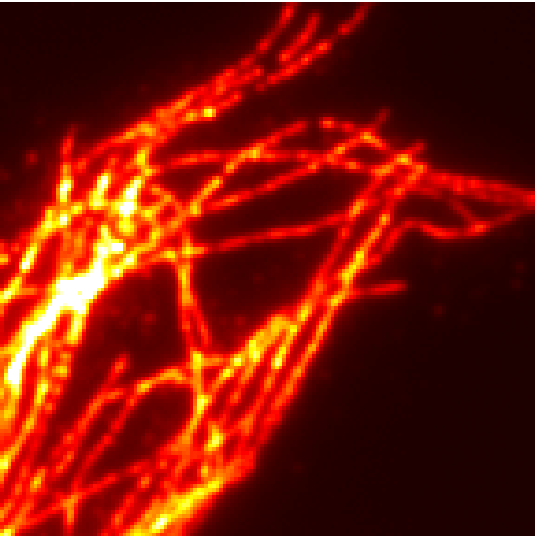}
     \end{subfigure}
     \hfill
      \begin{subfigure}[b]{0.08\textwidth}
         \centering
         \includegraphics[width=\textwidth]{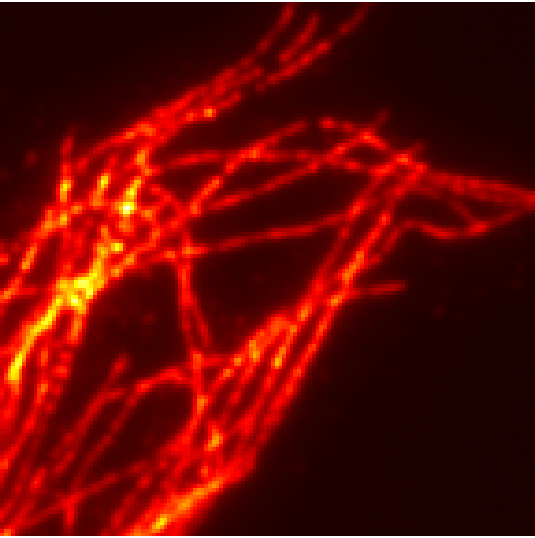}
     \end{subfigure}
     \hfill
      \begin{subfigure}[b]{0.08\textwidth}
         \centering
         \includegraphics[width=\textwidth]{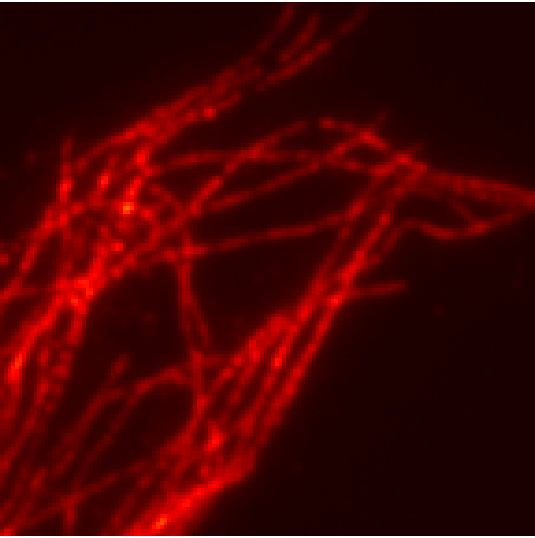}
     \end{subfigure}
     \hfill
      \begin{subfigure}[b]{0.08\textwidth}
         \centering
         \includegraphics[width=\textwidth]{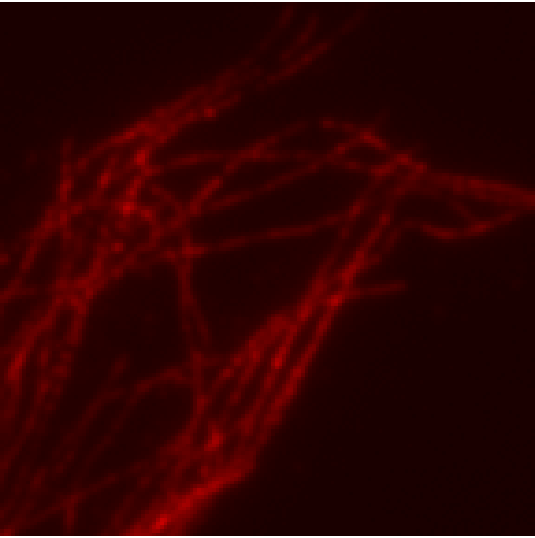}
     \end{subfigure}
     \hfill
      \begin{subfigure}[b]{0.018\textwidth}
         \centering
         \includegraphics[width=\textwidth]{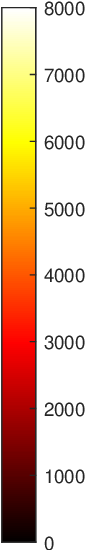}
     \end{subfigure}
     
    \vspace{0.1cm}
     
     \begin{subfigure}[b]{0.08\textwidth}
         \centering
         \includegraphics[width=\textwidth]{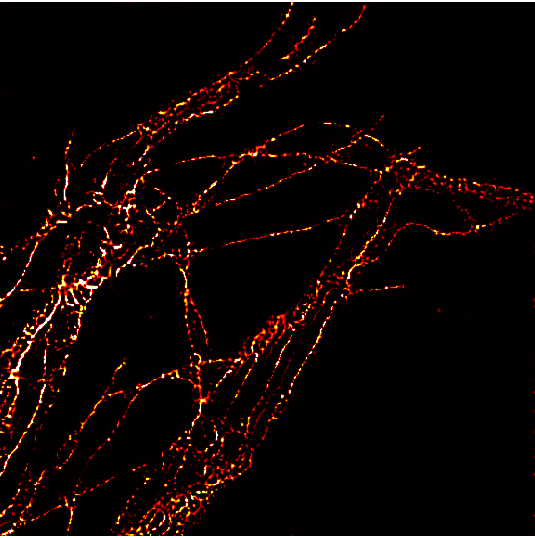}
         \caption*{$\alpha_1$}
     \end{subfigure}
     \hfill
      \begin{subfigure}[b]{0.08\textwidth}
         \centering
         \includegraphics[width=\textwidth]{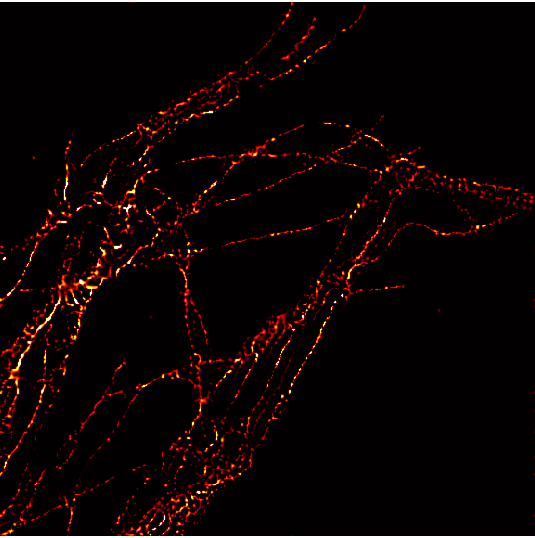}
         \caption*{$\alpha_2$}
     \end{subfigure}
     \hfill
      \begin{subfigure}[b]{0.08\textwidth}
         \centering
         \includegraphics[width=\textwidth]{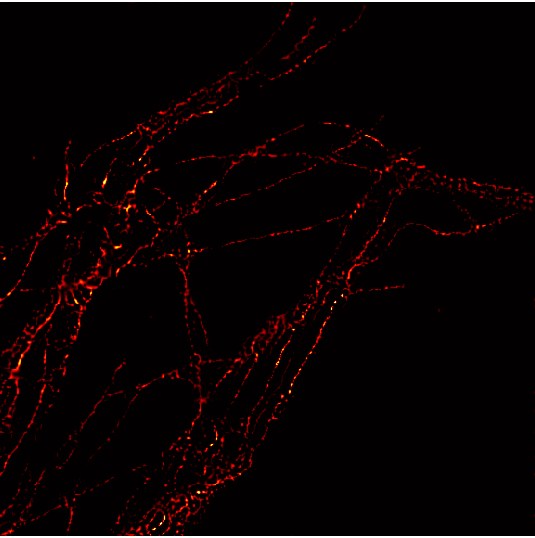}
         \caption*{$\alpha_3$}
     \end{subfigure}
     \hfill
      \begin{subfigure}[b]{0.08\textwidth}
         \centering
         \includegraphics[width=\textwidth]{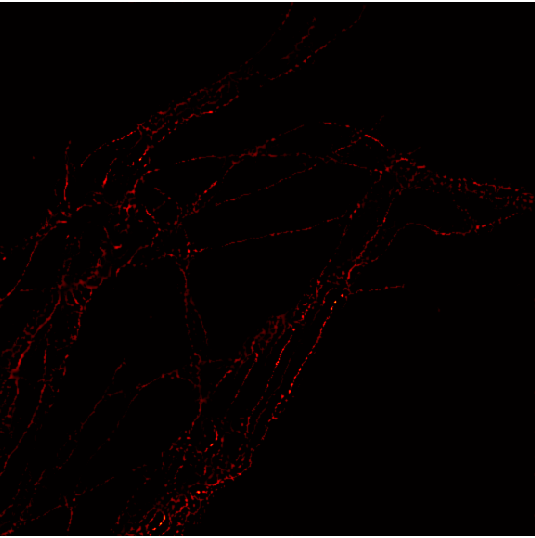}
         \caption*{$\alpha_4$}
     \end{subfigure}
     \hfill
      \begin{subfigure}[b]{0.018\textwidth}
         \centering
         \includegraphics[width=\textwidth]{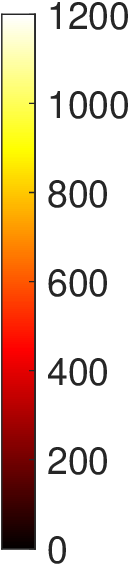}
         \caption*{}
     \end{subfigure}
     \caption{For each $\alpha_q, q=1,\ldots,4$ of the illumination beam: (first row) the temporal mean $\bar{\mathbf{g}}_{\alpha_q}$ of the stack, (second row) the 2D COL0RME results $\hat{\mathbf{x}}_{\alpha_q}$.}
     \label{real_data}
     
    \begin{subfigure}[b]{0.43\textwidth}
         \centering         \includegraphics[width=\textwidth]{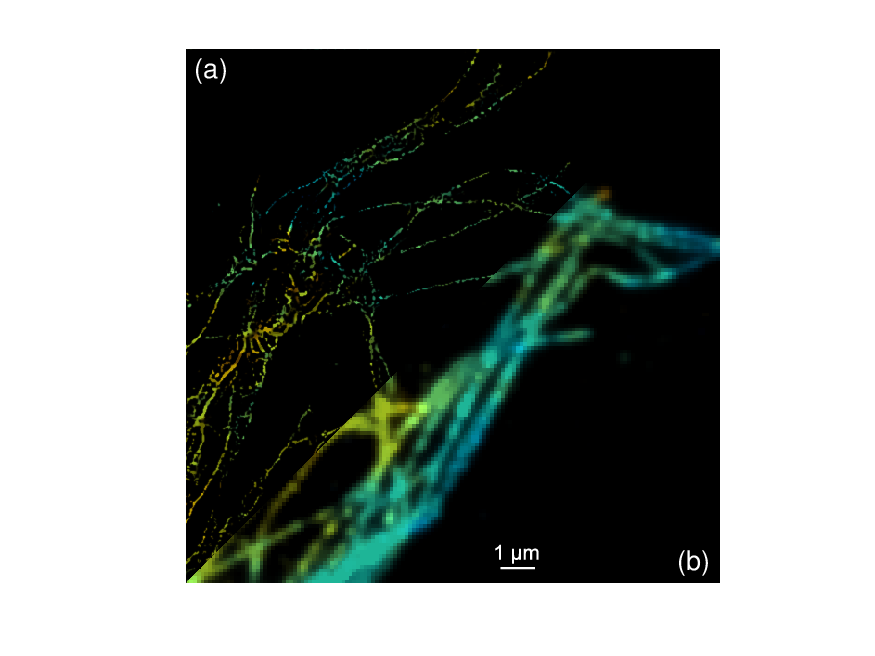}
     \end{subfigure}
      \begin{subfigure}[b]{0.04\textwidth}
         \centering
         \includegraphics[trim=190 0 190 0,clip,width=\textwidth]{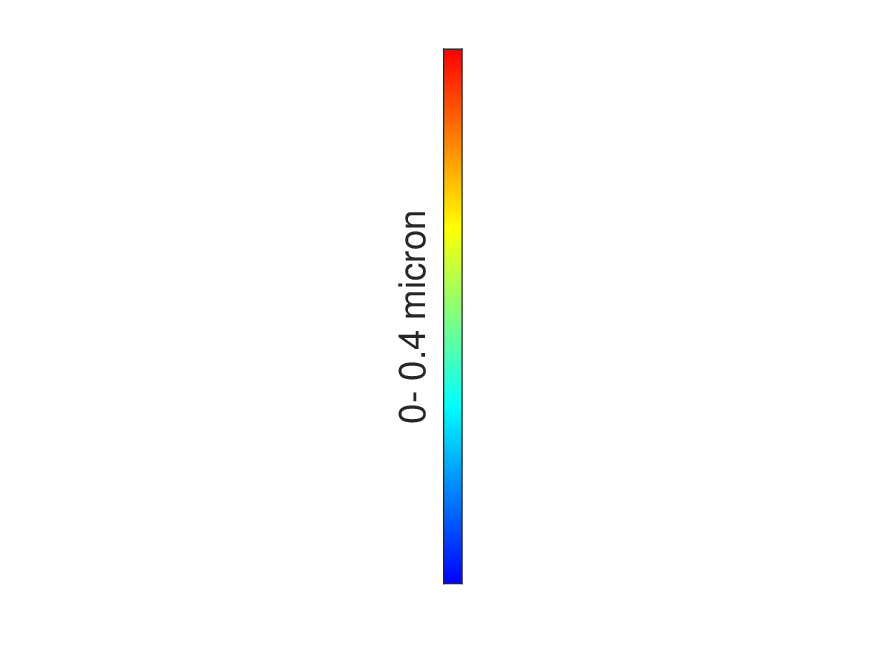}
     \end{subfigure}
   \vspace{-0.5\baselineskip}
     \caption{(a) Super-resolved 3D-COL0RME image, (b) Standard MA-TIRF reconstruction. Colour quantifies sample depth.}     
     \label{3D_COL0RME_real_data}
     \vspace{-0.6cm}
\end{figure}

In Figure \ref{real_data} (first row), the diffraction limited images are shown, together with the spatially super-resolved COL0RME images (second row). Finally, in Figure \ref{3D_COL0RME_real_data}, a comparison between the 3D reconstruction computed using 3D-COL0RME (left upper part) and the standard MA-TIRF approach with only background removal \cite{soubies_ISBI} (right lower part) is shown. 

\vspace{-0.1cm}
\footnotesize{\section{Compliance with ethical standards}
\label{sec:ethics}
\vspace{-0.3cm}
This work was conducted using biological data available in open access by EPFL SMLM datasets and data from cultured cells where no ethical questions are concerned.
}

\vspace{-0.1cm}
\footnotesize{\section{Acknowledgements}
\label{sec:acknowledgments}
\vspace{-0.3cm}
The authors are thankful to E. Soubies for providing code for MA-TIRF reconstruction. 
VS and LBF are supported by the 3IA
Côte d’Azur Investments (with reference number ANR-19-P3IA-0002).
LC acknowledges the EU H2020 RISE NoMADS grant 777826 and the SPLIN grant by GdR ISIS. Support for development of the microscope was received from IBiSA (Infrastructures en Biologie Santé et Agronomie) to the MICA microscopy platform.}

\vspace{-0.1cm}
\footnotesize
\bibliographystyle{IEEEbib}
\bibliography{strings,refs}

\begin{thebibliography}{10}

\bibitem{3D_STORM}
B.~Huang, W.~Wang, M.~Bates, and X.~Zhuang,
\newblock ``Three-dimensional super-resolution imaging by stochastic optical
  reconstruction microscopy,''
\newblock {\em Science}, vol. 319, no. 5864, pp. 810--813, 2008.

\bibitem{3DSTED}
B.~Harke, C.K. Ullal, J.~Keller, and S.W. Hell,
\newblock ``Three-dimensional nanoscopy of colloidal crystals,''
\newblock {\em Nano. Lett}, pp. 1309--1313, 2008.

\bibitem{iPALM}
G.~Shtengel, J.A. Galbraith, C.G. Galbraith, et~al.,
\newblock ``Interferometric fluorescent super-resolution microscopy resolves 3d
  cellular ultrastructure,''
\newblock {\em Proceedings of the National Academy of Sciences}, vol. 106, no.
  9, pp. 3125--3130, 2009.

\bibitem{3D_SIM}
M.G.L. Gustafsson, L.Shao, P.M. Carlton, et~al.,
\newblock ``Three-dimensional resolution doubling in wide-field fluorescence
  microscopy by structured illumination,''
\newblock {\em Biophysical Journal}, vol. 94, no. 12, pp. 4957--4970, 2008.

\bibitem{TIRF_axelrod81}
D.~Axelrod,
\newblock ``Cell-substrate contacts illuminated by total internal reflection
  fluorescence,''
\newblock {\em The Journal of cell biology}, vol. 89, no. 1, pp. 141—145,
  April 1981.

\bibitem{AXELROD2008169}
D.~Axelrod,
\newblock ``Chapter 7 total internal reflection fluorescence microscopy,''
\newblock in {\em Biophysical Tools for Biologists, Volume Two: In Vivo
  Techniques}, vol.~89 of {\em Methods in Cell Biology}, pp. 169--221. Academic
  Press, 2008.

\bibitem{stergiopoulou_ISBI}
V.~Stergiopoulou, J.~H. de~M.~Goulart, S.~Schaub, L.~Calatroni, and
  L.~Blanc-Féraud,
\newblock ``{COL0RME}: Covariance-based $\ell_0$ super-resolution microscopy
  with intensity estimation,''
\newblock in {\em 2021 IEEE 18th International Symposium on Biomedical Imaging
  (ISBI)}, 2021, pp. 349--352.

\bibitem{stergiopoulou_BioIm}
V.~Stergiopoulou, L.~Calatroni, J.~H. de~M. Goulart, S.~Schaub, and
  L.~Blanc-F{\'e}raud,
\newblock ``{COL0RME: Super-resolution microscopy based on sparse blinking
  fluorophore localization and intensity estimation},''
\newblock HAL preprint: \url{https://hal.archives-ouvertes.fr/hal-03320950},
  Aug. 2021.

\bibitem{soubies_journal}
E.~Soubies, A.~Radwanska, D.~Grall, L.~Blanc-F{\'e}raud, E.~Van
  Obberghen-Schilling, and S.~Schaub,
\newblock ``Nanometric axial resolution of fibronectin assembly units achieved
  with an efficient reconstruction approach for multi-angle-tirf microscopy,''
\newblock {\em {Scientific Reports}}, vol. 9, no. 1, Dec. 2019.

\bibitem{soubies_ISBI}
E.~Soubies, L.~Blanc-F{\'e}raud, S.~Schaub, and E.~Van Obberghen-Schilling,
\newblock ``{Improving 3D MA-TIRF Reconstruction with Deconvolution and
  Background Estimation},''
\newblock in {\em {IEEE International Symposium on Biomedical Imaging}},
  Venise, Italy, Apr. 2019.

\bibitem{CELO}
E.~Soubies, L.~Blanc-Féraud, and G.~Aubert,
\newblock ``A continuous exact $\ell_0$ penalty ({CEL0}) for least squares
  regularized problem,''
\newblock {\em SIAM Journal on Imaging Sciences, 8 (3)}, pp. 1607--1639, 2015.

\bibitem{chambolle_pock_2016}
A.~Chambolle and T.~Pock,
\newblock ``An introduction to continuous optimization for imaging,''
\newblock {\em Acta Numerica}, vol. 25, pp. 161–319, 2016.

\bibitem{DiscInvProb_Hansen}
P.C. Hansen,
\newblock {\em Discrete Inverse Problems: Insight and Algorithms},
\newblock Society for Industrial and Applied Mathematics, USA, 2010.

\bibitem{SOFItool}
A.~Girsault, T.~Lukes, A.~Sharipov, et~al.,
\newblock ``{SOFI} simulation tool: A software package for simulating and
  testing super-resolution optical fluctuation imaging,''
\newblock {\em PLOS ONE, 11 (9)}, , no. 9, pp. 1--13, 2016.

\end{thebibliography}

\end{document}